\documentclass[preprint,5p,times,authoryear]{elsarticle}
\usepackage{graphicx} 
\usepackage{url} 
\usepackage[hidelinks]{hyperref}
\usepackage{booktabs}
\usepackage{tabularx}
\usepackage{array}
\usepackage{subcaption}
\usepackage{color}
\usepackage{tikz}
\usepackage[most]{tcolorbox}

\newcommand{\MD}[1]{\textcolor{cyan}{#1}}

\newcommand*\WC[1]{%
\begin{tikzpicture}[baseline=(C.base)]
\node[draw,circle,inner sep=0.2pt](C) {#1};
\end{tikzpicture}}

\begin{document}
\begin{frontmatter}

\title{Forensic analysis of video data deletion and recovery \\in Honeywell surveillance file system}


\author[A1]{Jinhee Yoon}
\ead{lgrace22@g.skku.edu}

\author[A1]{Sungjae Hwang\corref{cor1}}
\ead{sungjaeh@skku.edu}
\cortext[cor1]{Corresponding author}

\affiliation[A1]{
    organization={Department of Computer Science and Engineering, Sungkyunkwan University},
    city={Suwon},
    country={Republic of Korea}
}


\begin{abstract}
Real-time video surveillance systems store recorded video using digital video recorders (DVRs) and network video recorders (NVRs).
To support continuous high-volume video storage, these devices employ specialized, nonstandard file systems that are often proprietary and undocumented.
This lack of documentation significantly increases the time and effort required for forensic analysis.
In this study, we analyze an undocumented proprietary file system used by Honeywell video surveillance devices---one that, to the best of our knowledge, has not been examined in prior work---and investigate its deletion mechanisms and demonstrate the feasibility of video recovery after deletion. We perform a file system analysis using a binary diffing technique and evaluate three deletion methods supported by the target device: \WC{1} formatting-based deletion, \WC{2} data expiration, and \WC{3} overwrite. For each method, we investigate changes in file system metadata and on-disk data structures and demonstrate the feasibility of video data recovery.
Our findings aim to support more efficient and accurate forensic investigations of Honeywell surveillance products and provide foundational insights into the analysis of proprietary file systems used in video recording devices.


%

\end{abstract}


\end{frontmatter}

\section{Introduction}
Real-time video surveillance systems have become essential monitoring infrastructure in modern society, serving various purposes such as crime prevention, incident reconstruction, and security management. These systems store video data through network video recorders (NVRs) and digital video recorders (DVRs), which record digital video streams using codecs such as MPEG-4 and H.264.
Video surveillance data provides an objective record of incidents and often serves as critical legal evidence, offering greater reliability than witness testimony. 
Accordingly, video surveillance forensics plays an important role in incident reconstruction, authenticity verification, and ensuring the reliability of digital evidence.

The scale and economic significance of video surveillance systems further show the importance of video surveillance forensics. 
According to a report by Mordor Intelligence~\citep{Mordor_NA_VideoSurveillance_2025}, 
the North American video surveillance market is valued at \$26.84 billion in 2025 and is projected to grow to \$50.48 billion by 2030, with a compound annual growth rate of 13.46\%. Within this relatively concentrated market, the top five suppliers account for a substantial portion of total revenue (21-32\%), and Honeywell is among these leading vendors.

Despite their widespread deployment, the internal file system structures used to store surveillance video data vary across manufacturers and are frequently undocumented or insufficiently studied. This lack of documentation poses significant challenges for digital forensic analysis and data recovery, particularly when deletion operations have occurred. 
While prior work has investigated the file system structures and forensic artifacts of surveillance devices from vendors such as Hikvision~\citep{hikvision2025} and Dahua~\citep{DahuaTechnology}, comparable analysis of Honeywell devices remains largely unexplored.

Previous studies have revealed the file system structure of Hikvision devices and enabled the extraction of video metadata and storage semantics~\citep{han2015_hikvision, sandeepa2018efficient}. Subsequent research has examined application-level interactions with Hikvision surveillance systems~\citep{dragonas2023iot2}, analyzed logs stored within their file systems~\citep{dragonas2023iot}, and extended forensic methodologies to Dahua surveillance devices~\citep{Dragonas2024_IoT_forensics_Dahua, rzayeva2025automated}. 
In contrast, to the best of our knowledge, no prior forensic study has examined the file system structure of Honeywell surveillance devices.

In this work, we analyze an undocumented proprietary file system used by Honeywell NVR devices and investigate its deletion mechanisms to examine the feasibility of video data recovery after deletion. To this end, we acquire disk images after applying three support deletion methods---formatting-based deletion, data expiration, and overwrite---and analyze changes in file system metadata and on-disk data structures. Using a binary diffing technique, we identify key file system structures and video management mechanisms related to deletion behavior. Based on this analysis, we demonstrate successful recovery of deleted video data across all three deletion methods. We believe that our findings provide practical insights for the forensic analysis of the surveillance video system.

\section{Related Work}

\noindent \textbf{File System Analysis.}
Studies on the file system analysis of surveillance systems have been continuously conducted in the digital forensics community. Because most DVR/NVR vendors employ proprietary and undocumented storage formats, prior research has necessarily focused on vendor-specific reverse engineering efforts. Representative systems—including Hikvision, Dahua, and in-vehicle cameras—have been analyzed, leading to diverse methodologies for identifying and understanding closed file system structures.
The first comprehensive structural analysis of the Hikvision DVR file system was presented by Han et al.~\citep{han2015_hikvision}. By reverse engineering the proprietary video viewer software, the authors identified metadata layouts, video storage formats, and indexing mechanisms. They further analyzed system initialization and overwriting processes, characterizing them as anti-forensic behaviors. This work established a foundational reference for subsequent forensic analyses of proprietary DVR file systems.
Similar efforts have been applied to other embedded surveillance platforms. Lee et al.~\citep{lee2023_rtos_fsidi} analyzed the proprietary file system used by in-vehicle built-in cameras operating under a real-time operating system (RTOS). Through chip-off extraction and reverse engineering of file system management software, the study demonstrated both file-level analysis and recovery of residual video frames from unallocated space.

\noindent \textbf{Video Recovery of Surveillance System.}
Several studies have focused on recovering video from proprietary DVR/NVR systems used by major manufacturers such as Dahua and Hikvision. Yang et al.~\citep{Yang2015_Dahua_Hikvision_Recovery} identified manufacturer-specific headers and footers that delimit H.264 video segments and demonstrated video recovery through carving techniques, even from damaged disks. However, because the underlying file system structures were not fully characterized, recovery efficiency was limited, as scanning was required across the entire disk image.
Building on the disclosure of the Hikvision file system structure, Sandeepa et al.~\citep{sandeepa2018efficient} proposed an automated recovery approach that leveraged known metadata regions to identify valid video data areas. While effective when metadata was present, the approach was unable to recover video in scenarios where metadata had been deleted or corrupted.
More recently, Rzayeva et al.~\citep{rzayeva2025automated} proposed an automated forensic framework for video recovery from Hikvision and Dahua DVR/NVR systems. Their method identifies the manufacturer from disk images, extracts video frames using vendor-specific signatures, and reconstructs timelines using adaptive temporal interval analysis. This work demonstrates improved recovery accuracy and temporal consistency compared to conventional carving techniques and commercial tools.

\noindent \textbf{Logs and Behavioral Artifacts of Surveillance System.}
While early research primarily focused on video recovery, recent studies have expanded the scope of surveillance forensics to include system logs and behavioral artifacts. Dragonas et al.~\citep{dragonas2023iot} analyzed log record structures in Hikvision devices, demonstrating that logs can serve as independent forensic evidence of user access and system events. The authors also developed an automated log analysis tool, expanding forensic capabilities beyond video-centric analysis.
Similarly, Dragonas et al.~\citep{Dragonas2024_IoT_forensics_Dahua} conducted a systematic analysis of log artifacts across storage media, file systems, and internal memory in Dahua surveillance systems. Their experiments correlated logs with user actions and evaluated the impact of anti-forensic operations such as formatting and log deletion, further advancing behavior-based forensic analysis.
Complementary work by Dragonas et al.~\citep{dragonas2023iot2} extended surveillance forensics beyond DVR/NVR hardware by analyzing mobile applications used to control Hikvision systems. By examining Android and iOS app artifacts, the authors identified evidence of user authentication, system access, and viewing behaviors, and integrated their findings into ALEAPP~\citep{brignoni_aleapp} and iLEAPP~\citep{brignoni_ileapp}.

Prior research shows that reverse engineering and structural analysis are effective for proprietary surveillance file systems, but vendor-specific designs limit cross-applicability. This study addresses this gap by presenting the first systematic analysis of the undocumented Honeywell NVR file system, examining its deletion mechanisms and demonstrating video recovery.

\begin{figure*}[t]
  \centering
  \includegraphics[width=\linewidth]{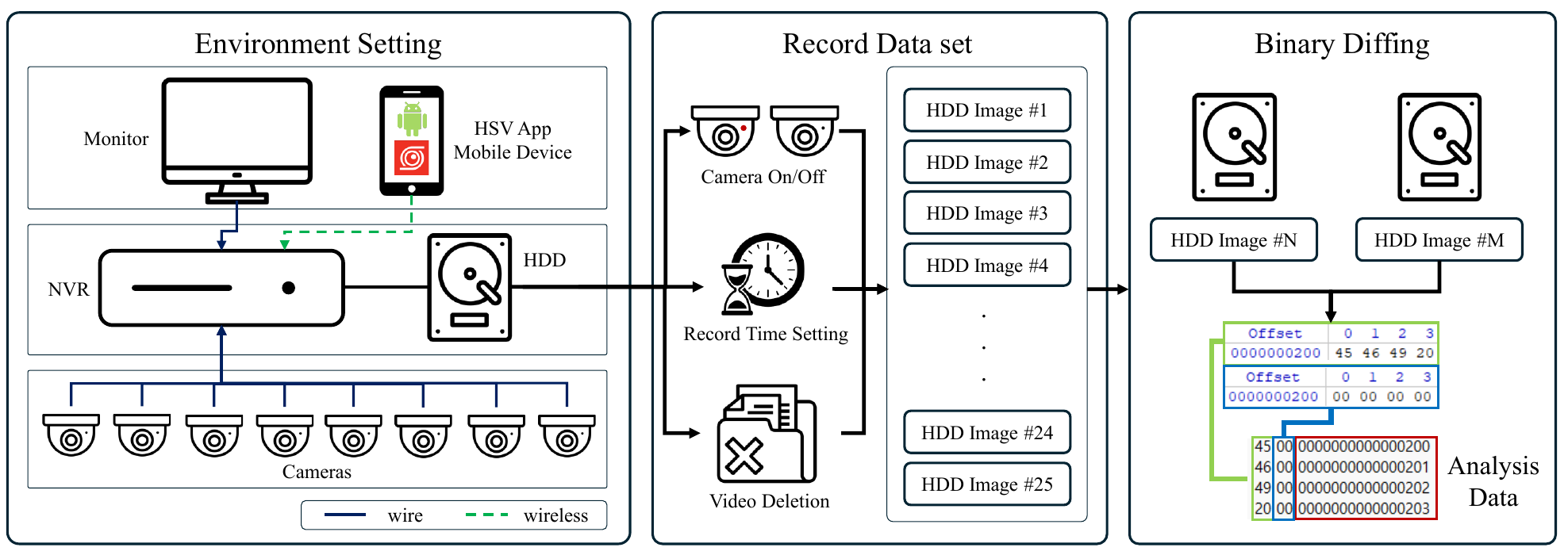}
  \caption{Research workflow: environment setup, dataset acquisition during recording and deletion operations, and binary diffing analysis.}
  \label{fig:Research_Workflow}
\end{figure*}
\section{Background}
\subsection{H.264 NAL Units}
H.264 (also known as AVC, Advanced Video Coding) is one of the most widely used video compression standards in surveillance devices such as cameras, NVRs, and DVRs~\citep{KWON2006186}. It enables long-term recording under limited storage constraints through high compression efficiency and multiple profiles. A core component of H.264 is the Network Abstraction Layer (NAL) unit, which encapsulates video data for network transmission and storage.
NAL units are categorized as IDR (Instantaneous Decoder Refresh), non-IDR, and parameter sets, including the Sequence Parameter Set (SPS) and Picture Parameter Set (PPS). IDR NAL units correspond to I-frames and can be decoded independently, whereas non-IDR NAL units (e.g., P- and B-frames) are decoded by referencing I-frames. Moreover, both IDR and non-IDR NAL units contain video data and require the associated SPS and PPS for decoding~\citep{RFC6184}.

Understanding NAL unit structure is important for file system analysis of video storage devices for two reasons. First, when raw storage consists of contiguous H.264 streams, distinctive NAL header patterns serve as reliable markers for identifying data regions during fragmentation or deletion analysis. Second, some vendors store data on a per–NAL-unit basis and prepend proprietary headers before standard NAL headers; analyzing these headers can reveal additional information about the storage layout and video data.

\subsection{Byte-level Binary Diffing}
Binary diffing based on byte-level comparison is a technique that identifies differences by comparing two or more binary datasets at the byte level~\citep{nist800168}. It is widely used in digital forensics and reverse engineering to identify data regions modified by specific operations, and is particularly effective for analyzing proprietary file systems with undocumented structures. By comparing disk images acquired from the same device initialized to an identical state and subjected to different operations, analysts can directly observe how each operation affects the internal file system layout. Even without prior knowledge of the file system, comparison with a raw-formatted disk image enables rapid identification of regions containing meaningful data.

In this study, we employ binary diffing to analyze the file system structure of Honeywell NVR devices. This approach allows us to identify the overall file system layout and how three deletion methods---formatting-based deletion, data expiration, and overwrite---affect internal data structures.



\begin{table*}[t]
\centering
\footnotesize
\caption{HDD images used in the experiments. Sample denotes video recording, while f-deletion, e-deletion, and o-deletion denote video deletion via formatting, expiration, and overwriting, respectively. Formatting refers to initializing a disk.}
\label{tab:nvr_summary}
\begin{tabularx}{\textwidth}{l r X}
\toprule
\textbf{Image \#} & \textbf{Name} & \textbf{Description} \\
\midrule
01 & row format & The 160GB hard disk image completely overwritten with zeros.  \\
02 & formatting (1) & The 160GB hard disk image immediately after the first formatting. \\
03 & sample (1.1) & The 160GB hard disk image includes 24 hour recorded video from 1 camera using Image 02. \\
04 & sample (1.2) & The 160GB hard disk image includes 1 hour recorded video from 1 camera using Image 02. \\
05 & formatting (2) & The 160GB hard disk image immediately after the second formatting. \\
06 & sample (2.1) & The 160GB hard disk image recorded over 5 minutes from 2 cameras using Image 05. \\
07 & sample (2.2) & The 160GB hard disk image perform device access only using Image 06 without capturing video. \\
08 & f-deletion (2.1.1) & The 160GB hard disk image with formatting deletion performed using Image 06. \\
09 & formatting (3) & The 160GB hard disk image immediately after the third formatting. \\
10 & sample (3.1) & The 160GB hard disk image recorded using Image 09, capturing 10 separate 5-minute recordings from 1 camera. \\
11 & sample (3.2) & The 160GB hard disk image recorded once using image 09, with each camera recording for 50 minutes from 1 camera. \\
12 & sample (3.3) & The 160GB hard disk image recorded over 4 hours and 30 minutes using 1 camera with Image 09. \\
13 & sample (3.4) & The 160GB hard disk image recorded for 5 minutes using 2 cameras with Image 09. \\
14 & sample (3.5) & The 160GB hard disk image recorded for 5 minutes using 8 cameras with Image 09. \\
15 & e-deletion (3.5.1) & The 160GB hard disk image for Image 14 was deleted due to expiration of its retention period(1 day). \\
16 & f-deletion (3.5.2) & The 160GB hard disk for Image 14 was deleted via formatting deletion. \\
17 & sample (3.5.3) & The 160GB hard disk image recorded for 5 minutes using 8 cameras with image 16. \\
18 & sample (3.6) & The 160GB hard disk image using Image 09, captured with 8 cameras without deletion until disk space ran out. \\
19 & e-deletion (3.6.1) & The 160GB hard disk image for Image 18 was deleted due to expiration of its retention period(4 days). \\
20 & o-deletion (3.6.2) & The 160GB hard disk image created by overwriting data 20 minutes from image 18 using 8 cameras. \\
21 & f-deletion (3.6.3) & The 160GB hard disk image in Image 18 was deleted via formatting deletion. \\
22 & sample (3.7) & The 160GB hard disk image using Image 09, recorded only the main stream with 8 cameras.\\
001 & row format & The 250GB hard disk image completely overwritten with zeros. \\
002 & formatting (1) & The 250GB hard disk image immediately after the first formatting. \\
003 & sample (1.1) & The 250GB hard disk image with video recorded using Image 002. \\
\bottomrule
\end{tabularx}
\end{table*}
\section{Methodology}
\autoref{fig:Research_Workflow} illustrates the workflow of our study, which consists of three stages: \WC{1} environment setup for operating Honeywell surveillance devices; \WC{2} dataset acquisition, in which deletion operations are performed and file system images are dumped; and \WC{3} binary diffing analysis of the dumped images to investigate the file system structure and its handling of deletion operations. Each stage is described in detail in the following subsections.

\subsection{Environment Setup}
We constructed an experimental environment using a Honeywell NVR (model HN35080200), a monitor, a mobile device running the HSV remote-access application, eight HN40E-2030I cameras, and Seagate hard disks with capacities of 160~GB and 250~GB. All cameras were connected to the NVR via Power over Ethernet (PoE) and configured with identical settings for continuous recording starting simultaneously. The monitor and mobile device were used to control the NVR and view recorded video. This device supports H.264/265, but the experiment was conducted using the H.264 codec. The environment was configured to closely reflect typical real-world surveillance deployments.
 
\subsection{Dataset Acquisition}
To analyze the file system structure of the Honeywell NVR and the effects of different deletion methods, we collected disk images from two hard disks with capacities of 160~GB and 250~GB under multiple experimental conditions. Disk images were acquired using FTK Imager~\citep{ftk_imager}, and their descriptions are summarized in \autoref{tab:nvr_summary}.
As a baseline, we first created a disk image filled entirely with 0x00 values (Image \#01). We then generated experimental datasets by varying three parameters: \WC{1} the number of recording channels, \WC{2} the recording duration, and \WC{3} the data deletion methods. 

Honeywell NVR supports three deletion methods: \textit{formatting-based} deletion, which removes all stored video data by reformatting the disk; \textit{data expiration}, which deletes video data whose retention period has elapsed; and \textit{overwrite}, which replaces the oldest data when the disk becomes full. Disk images were acquired after each operation to capture file system changes resulting from deletion. Accordingly, we collected images corresponding to formatting-based deletion (Images \#08, \#16, and \#21), data expiration (Images \#14 and \#19), and overwrite (Image \#20).

For the 250 GB disk, as a comparison set for conducting the same analysis, an initial 0x00-filled image (Image \#001), an image acquired immediately after formatting (Image \#002), and an image obtained after recording for 5 minutes following formatting (Image \#003) were generated and used to analyze file system structures with respect to capacity differences. A 250 GB disk image was used to evaluate scalability. While partition sizes (e.g., the video data region) vary depending on capacity, the overall structure and deletion behavior remain consistent.

\subsection{Binary Diffing}
We implemented a binary diffing tool to efficiently identify differences across multiple disk images. The tool compares data in block-sized units to support large-capacity storage devices and performs byte-level comparisons to detect modified regions 
resulting from operations such as deletion, formatting, and overwriting. For each difference, both the absolute byte offset and the corresponding relative offset within the analyzed partition were recorded, enabling precise localization of file system structures for further analysis. The binary diffing tool is publicly available~\citep{honeywell_nvr_tools}. 
\begin{figure}[t]
  \centering
  \includegraphics[width=\linewidth]{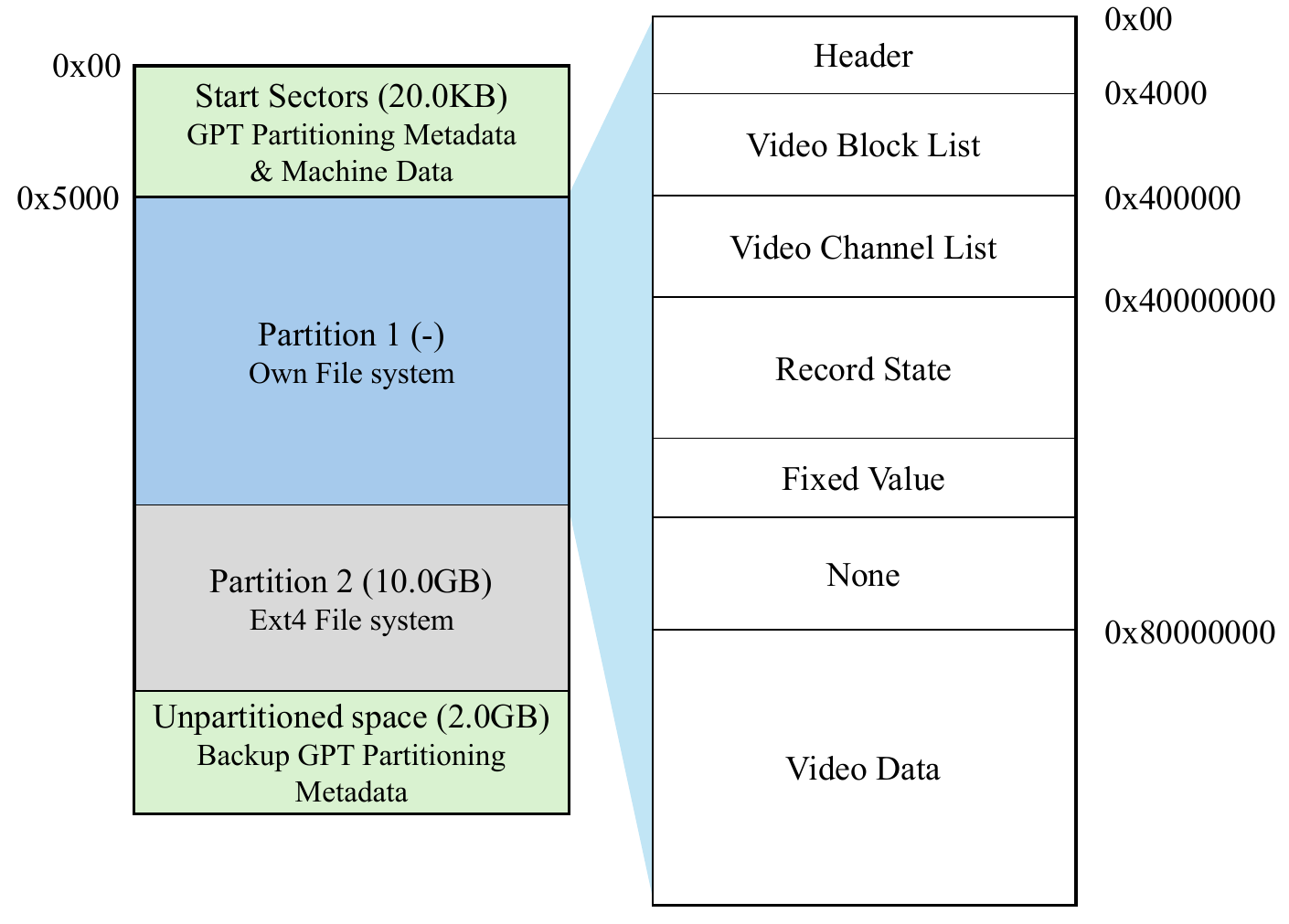}
  \caption{Overall file system layout of the Honeywell NVR.}
  \label{fig:Honeywell NVR File System Structure}
\end{figure}

\section{Honeywell NVR File System}
This section describes the file system used by the Honeywell NVR, as derived from our binary diffing analysis. 

\subsection{Overall File System Layout}
\autoref{fig:Honeywell NVR File System Structure} illustrates the overall file system layout of the Honeywell NVR. The device uses a GUID Partition Table (GPT) sheme~\citep{UEFI_GPT_2_10}, consisting of two main partitions, along with start sectors and unpartitioned space.

The first 20 KB at the beginning of the disk, referred to as the start sectors, contain essential metadata that form the GPT structure, including the Protective MBR, the Primary GPT Header, and the Partition Entry Array, as well as device-specific configuration and identification information. This region plays a critical role in enabling the NVR system to recognize and operate the disk. 
Immediately following the start sectors, Partition 1 is a large, variably sized region that serves as the primary storage area for video data and is the main focus of our analysis. It contains Honeywell's proprietary data storage scheme, including video data blocks and indexing structures, and is where byte changes due to deletion and overwriting primarily occur.
Partition 2 is a fixed 10 GB region configured with the Ext4 file system. This partition is considered to store system-related files and configuration information necessary for device operation. It is not directly related to video data, and because it uses a well-documented file system, its importance for analysis is relatively low. 
At the end of the disk, there exists 2 GB of unpartitioned space, which includes the Secondary GPT Header and the Secondary Partition Entry Array. These are standard backup components of the GPT structure and serve as a safeguard for restoring partition information in the event of disk damage.

Our structural analysis shows that the Honeywell NVR disk combines a proprietary video storage partition with a standard file system–based system partition under a GPT layout. Our analysis focuses on Partition~1, which stores video data and is not publicly documented, as well as the start sectors and unpartitioned space containing fundamental disk information. Each region is described in detail in the following subsections, and partition names in each region are assigned arbitrarily based on our analysis.


%


\subsection{Start Sectors}
\begin{figure}[t]
  \centering
  \includegraphics[width=0.8\linewidth]{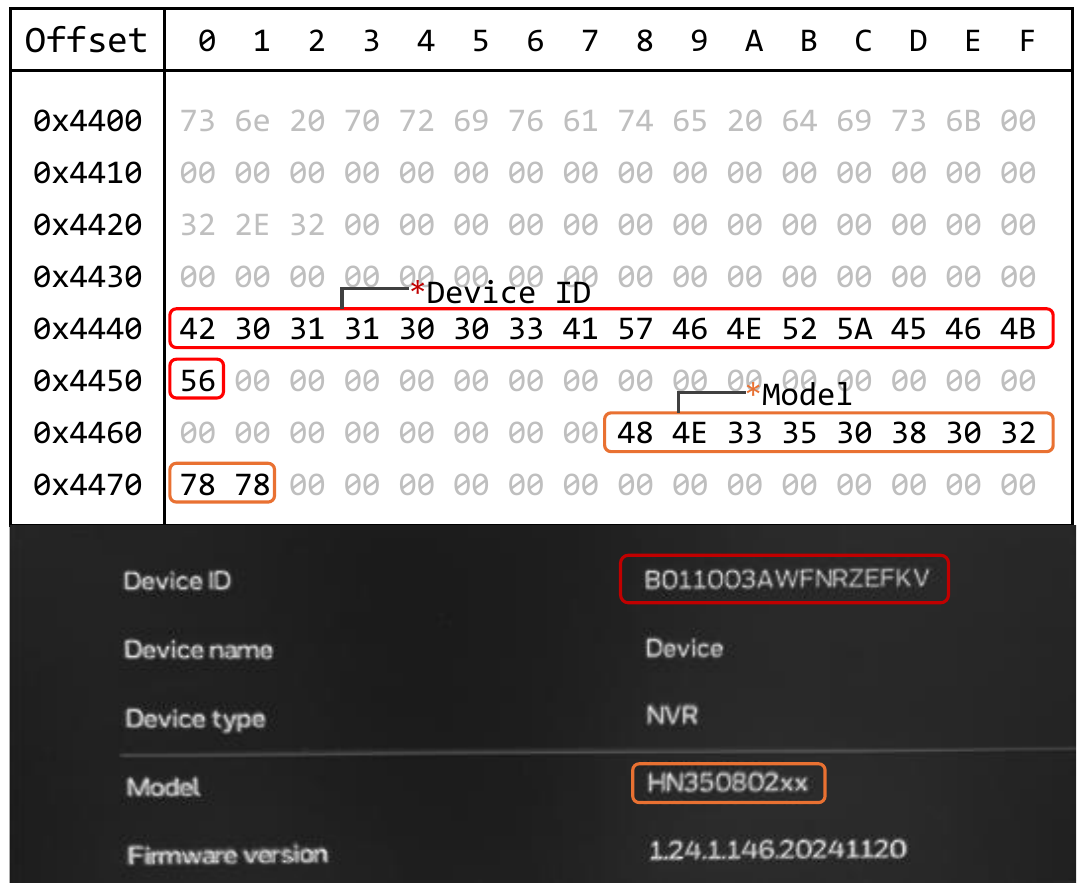}
  \caption{Machine Data region at Sector~34 within the start sectors, containing Honeywell NVR device identification information.}
  \label{fig:Machine Data}
\end{figure}

Honeywell NVR file system begins with the start sectors. This region consists of 40 sectors (approximately 20 KB) and stores metadata required for GPT-based disk operation as well as device-specific configuration information. Sector 0 contains the Protective MBR, which enables disk recognition by systems that do not support GPT, and Sector 1 contains the Primary GPT Header, which describes the disk layout and partition configuration. Sector 2 stores the Partition Entry Array, which defines the starting locations, sizes, and GUIDs of each partition, confirming that the disk is divided into two main partitions. Sectors 3 through 33 form a None area that is not used for any specific purpose and is considered either unused or reserved space. As shown in Figure \ref{fig:Machine Data}, Sector 34 contains a Machine Data region that records device-specific information such as the Honeywell NVR device ID and model name. This information is forensically significant as it identifies the device where the disk was installed. Sectors 35 through 39 again consist of a None area and contain no meaningful data.

\begin{figure}[t]
  \centering
  \includegraphics[width=0.9\linewidth]{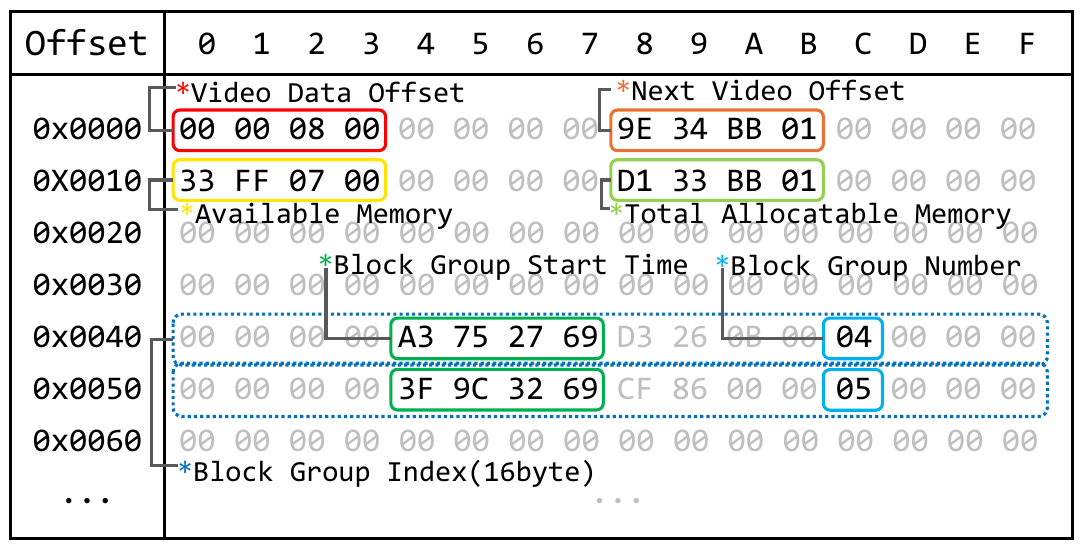}
  \caption{Structure of the Header area in Partition 1.}
  \label{fig:Header}
\end{figure}

\subsection{Unpartitioned Space}
At the end of the Honeywell NVR file system, 2.0 GB of unpartitioned space stores the Secondary GPT metadata in accordance with the GPT standard. This region contains no data other than the backup metadata. Sector 0 of this region stores the Secondary GPT Partition Entry Array, which backs up the Primary Partition Entry Array located in the start sectors and enables recovery of partition information if the primary structure is damaged. Sectors 1-31 and 33-4,194,303 are reserved areas that contain no recorded data. Sector 32 stores the Secondary GPT Header, which contains essential metadata for restoring GPT components, including disk layout, partition ranges, and CRC32 checksum information. The Secondary GPT Header ensures full recovery of partition information in the event of Primary GPT Header corruption.

\subsection{Partition 1}
Partition 1 of the Honeywell NVR storage device is a large, dynamically allocated video storage region. Our analysis confirmed that data are stored in little-endian format and organized into multiple function-specific areas: the Header, Video Block List, Video Channel List, Record State, Fixed Value, and Video Data regions. Each area is described below.


\begin{figure}[t]
  \centering
  \includegraphics[width=0.9\linewidth]{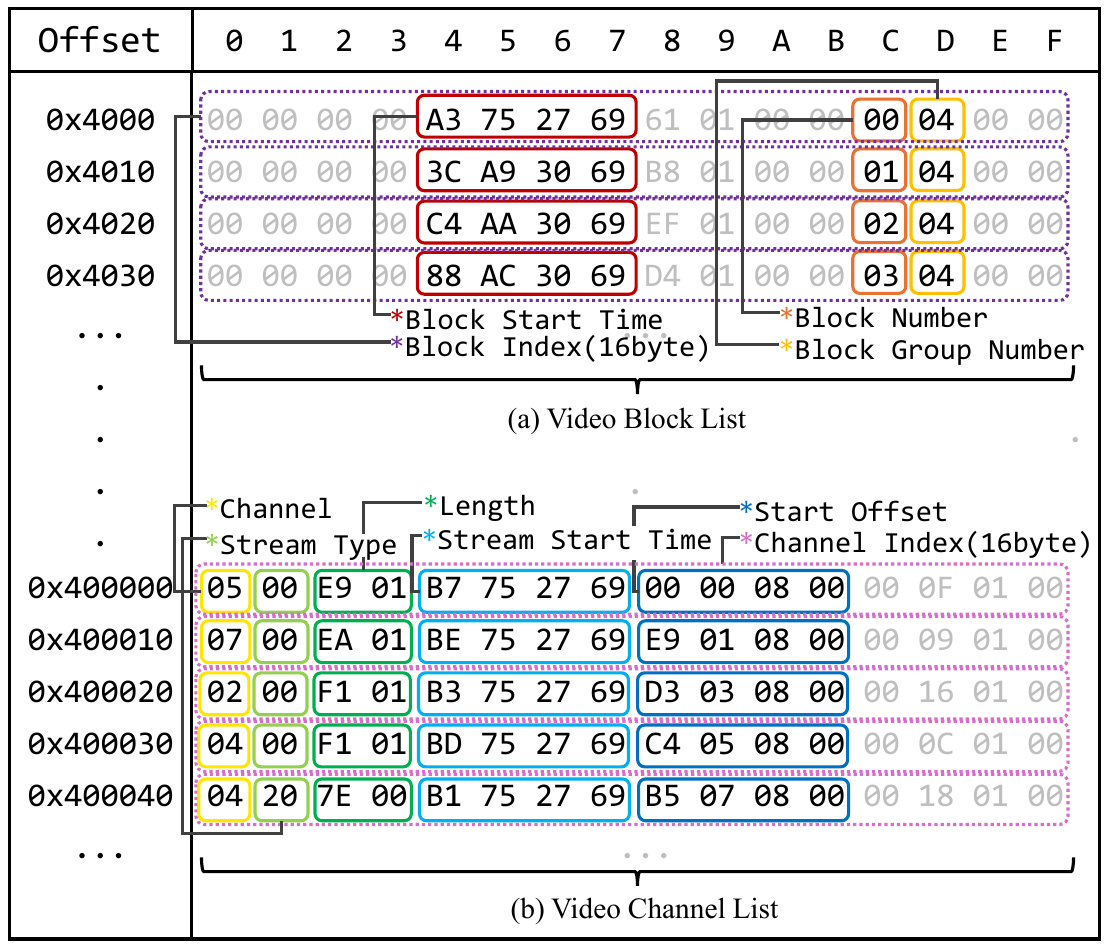}
  \caption{Structure of the (a) Video Block List, (b) Video Channel List areas in Partition 1.}
  \label{fig:BlockFrame}
\end{figure}

\subsubsection{Header}
\autoref{fig:Header} shows the header structure of Partition 1, which occupies range from 0x00 to 0x3FFF. The 4 bytes at offsets 0x00–0x03 specify the starting offset of the video data, rounded to the nearest thousand at the third digit; thus, video data begin at offset 0x80000000. The 4 bytes at offsets 0x08–0x0B indicate the offset for the next video write (0x01BB349E000 in \autoref{fig:Header}). Offsets 0x10–0x13 represent the available memory, while offsets 0x18–0x1B indicate the total allocatable memory. Notably, this total does not correspond to the full physical storage capacity: when the allocatable region is filled, the device reports a ``disk full'' state, although additional physical space remains. Deleted video data are treated as non-existent and do not consume allocatable storage.




A 16 byte block group index exists starting at offset 0x40. This region allows identification of the recording start time and block group. The 16 byte Block Group Index with this structure is continuously added as the video is recorded. The 4 byte value located at offsets 0x44–0x47 denotes the start time of the block group located at offset 0x4C. For example, the value 0x692775A3 corresponds to the Unix timestamp of November 26, 2025, 21:48:19. 

The timestamp field indicating the block group start time represents the time when the first video was captured for each block group. Data prior to the capture time cannot be viewed on the NVR device. If video data containing time stamps prior to the block start time remains in the video data field, it is highly likely that deletion attempts have been made.

\subsubsection{Video Block List}
The Video Block List occupies offsets 0x40000 to x3FFFFF of Partition 1. As shown in \autoref{fig:BlockFrame}(a), each entry is a fixed 16 byte block index. The first 4 bytes are reserved (set to 0x00), the next 4 bytes store the block start time, the 12th byte indicates the block number within a block group, and the 13th byte denotes the block group number. Entries are recorded sequentially, with each block group containing 256 blocks (0x00–0xFF). When a block group is full, subsequent entries are assigned to the next group. 

These repeated records are key artifacts for correlating block metadata with video data, enabling reconstruction of the file system structure and assessment of video recovery feasibility.


\subsubsection{Video Channel List}
Offsets 0x400000 to 0x3FFFFFFF of Partition 1 contain the Video Channel List for each camera channel (see \autoref{fig:BlockFrame}(b)). Each 16 byte Channel Index entry consists of a 1 byte channel identifier, a 1 byte stream type (0x00 for main, 0x20 for sub), a 2 byte frame length, a 4 byte frame start time, and a 4 byte frame start offset. Stream entries are not recorded in a fixed order. This list enables the device to locate channel data within the Video Data region and efficiently access frames during playback. Frame length and offset values are rounded at the third digit; for example, the little-endian value E9 01 corresponds to 0x01E9000, and 00 00 08 00 corresponds to 0x08000000.

Because the presence of this metadata varies with the deletion method, it is critical for identifying deletion causes and assessing video.


\begin{figure}[t]
  \centering
  \includegraphics[width=0.9\linewidth]{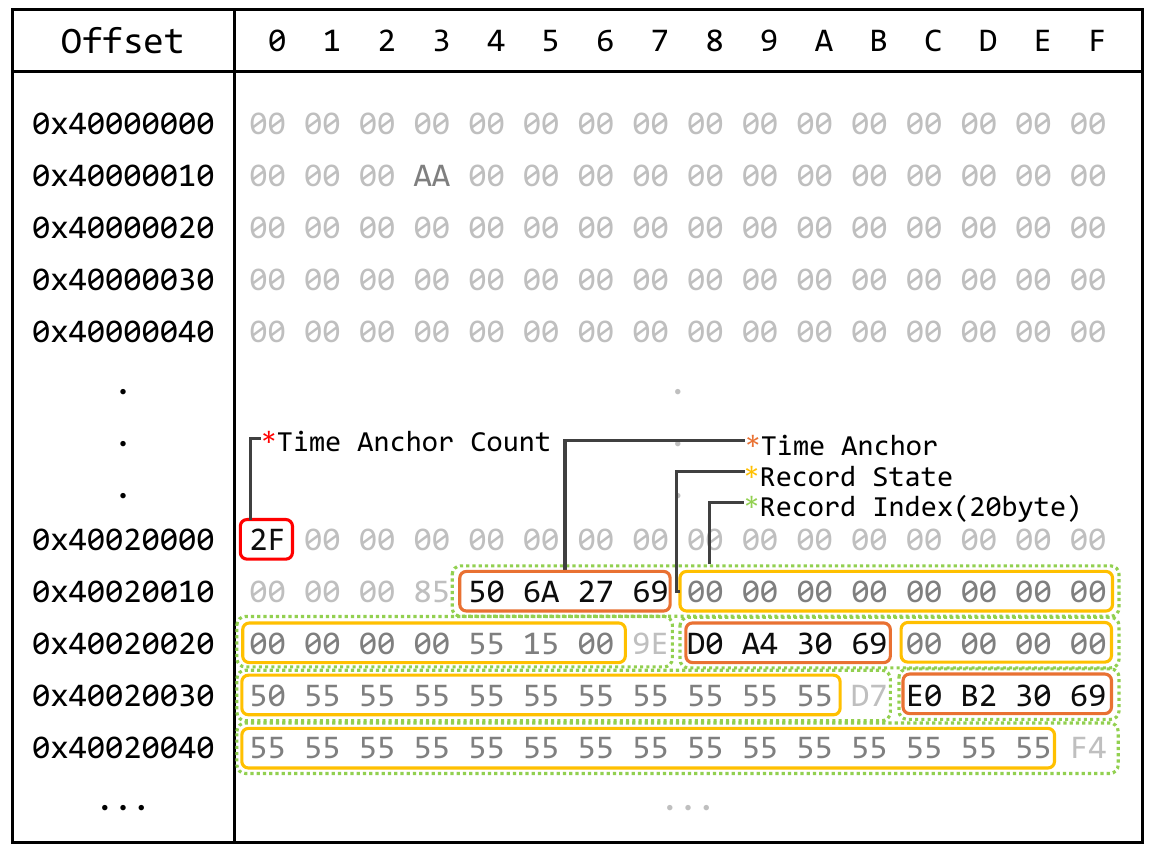}
  \caption{Structure of the Record State area in Partition 1.}
  \label{fig:Rstate}
\end{figure}

\subsubsection{Record State}
Offsets 0x40000000 to 0x4011FFFF of Partition 1 store the video record state for each channel. This region is divided into (number of channels + 1) subregions of 0x20000 bytes each. As shown in \autoref{fig:Rstate}, the first subregion (0x40000000 to 0x4001FFFF) is unassigned, channel 1 starts in the second subregion (0x40020000 to 0x4003FFFF), and channel 8 occupies the ninth. This layout scales dynamically as additional camera channels are added.

This region appears to store per-hour recording metadata. The first byte indicates the number of time anchors, corresponding to full-hour timestamps within the recording period. Each anchor is represented by a 20 byte Record Index entry: the first 4 bytes store the full-hour timestamp, followed by 15 bytes encoding the recording status for the corresponding hours.
For example, recordings from November 26, 2025, 21:49–21:55 and from December 3, 2025, 21:18 to December 5, 2025, 18:22 yield 47 full-hour time anchors. We observed timestamps increase in one-hour increments, indicating that this area functions as an index table recording per-hour recording status.

\begin{figure}[t]
  \centering
  \includegraphics[width=0.9\linewidth]{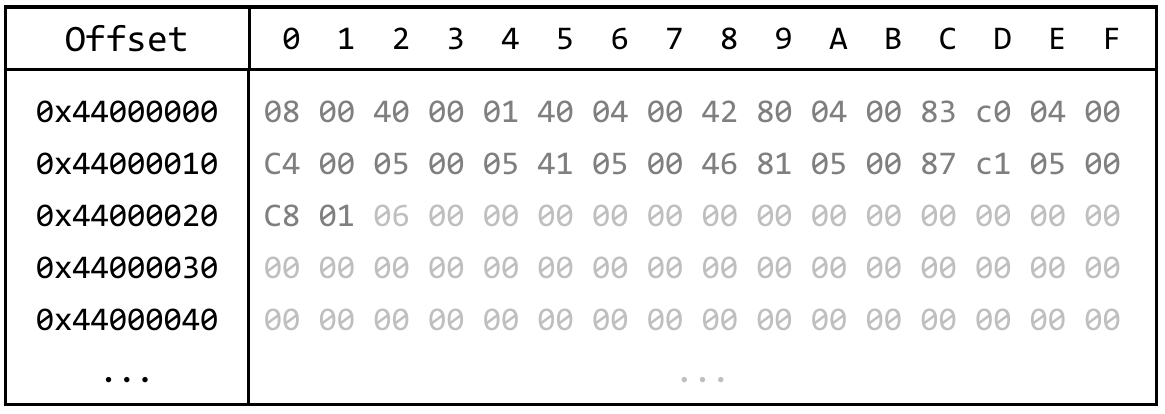}
  \caption{Data observed in the Fixed Value region.}
  \label{fig:FixData}
\end{figure}

\begin{figure}[t]
  \centering
  \includegraphics[width=0.9\linewidth]{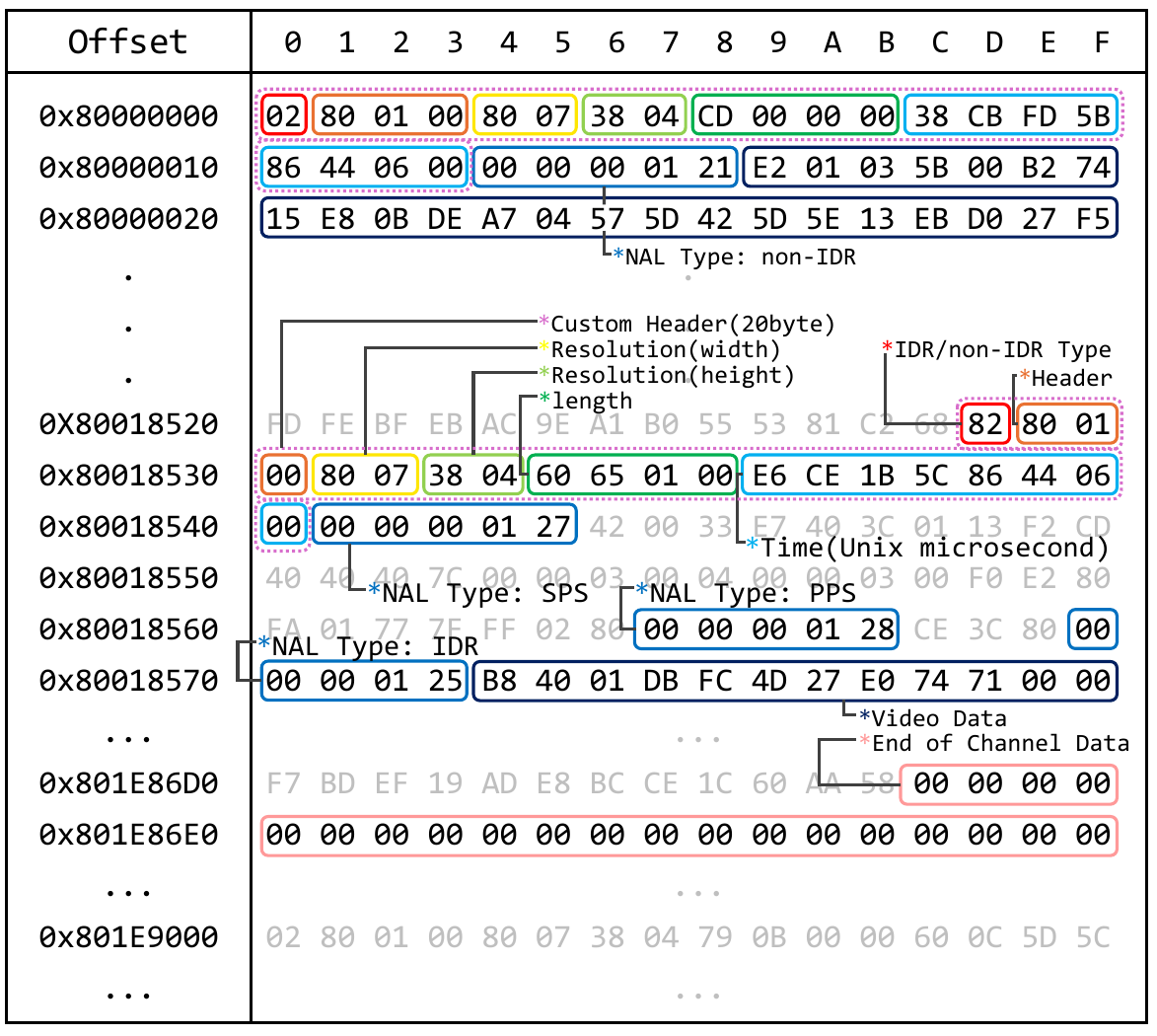}
  \caption{Structure of the Video Data area in Partition 1.}
  \label{fig:videoData}
\end{figure}

\subsubsection{Fixed Value}
The region following the Record State area contains fixed values (see \autoref{fig:FixData}) that remain unchanged and are unrelated to recording behavior, suggesting that they are device- or firmware-specific. As this region is not involved in video recording or recovery of deleted data, we did not analyze it further.


\begin{figure*}[t]
  \centering
  \includegraphics[width=0.96\linewidth]{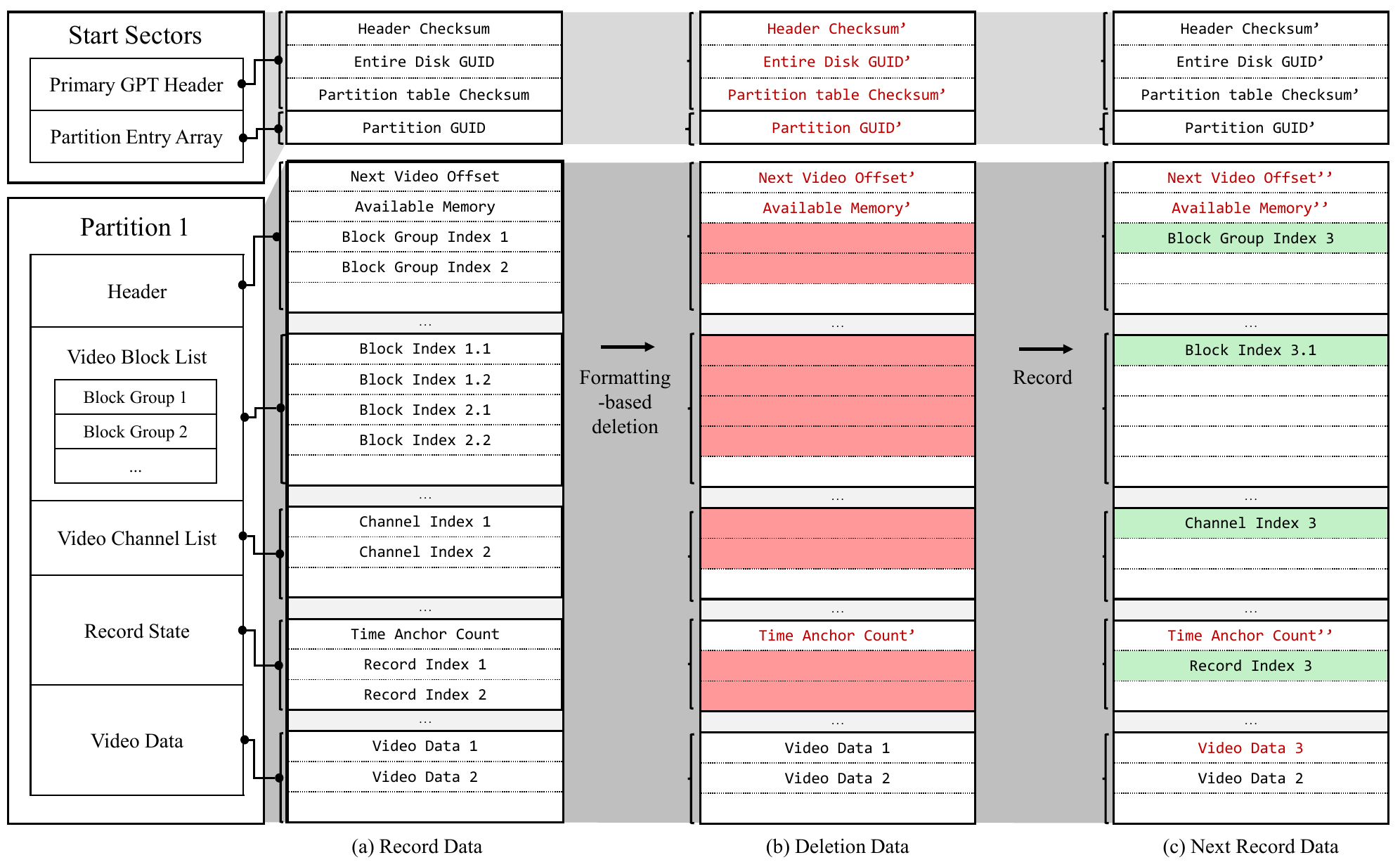}
  \caption{Changes in Start Sectors and Partition 1 caused by formatting-based deletion.}
  \label{fig:FDeletion}
\end{figure*}

\subsubsection{Video Data}
\label{videodata}
From offset 0x80000000 of Partition 1 to the end of the partition is the region in which video data are stored. In this region, video data are recorded in the form of NAL units using the H.264 codec. \autoref{fig:videoData} illustrates this video data region; in this file system, a 20 byte Custom Header precedes each NAL unit. In the Custom Header, the first byte indicates the frame type, with a value of 0x82 representing an IDR frame and 0x02 representing a non-IDR frame. This is followed by a fixed 3 byte value of 80 01 00. The next 4 bytes represent the video resolution, with 2 bytes each for width and height. Subsequently, a 4 byte value indicating the length of the NAL unit is recorded, followed by an 8 byte Unix microsecond timestamp. 

After the 20 byte Custom Header, 6 bytes consisting of the NAL unit start code and NAL unit header follow, after which the video data continue. Video Data is stored per channel as identified in the Video Channel List, and its length corresponds to the value specified in the Video Channel List. Video data for each channel is separated by a 20 byte value called End of Channel Data consisting entirely of 00s. A dummy value is stored in the gap between the actual data length and the rounded data length. In \autoref{fig:videoData}, a non-IDR and IDR frame of the main stream with a resolution of 1920×1080 can be observed. IDR frame has a length of 0x016560 bytes, and the timestamp value 0x000644865C1BCEE6 indicates that it is a I-frame recorded at 21:48:41.896 on November 26, 2025. 

This region stores the most critical video data in a real-time video surveillance environment and is therefore the first area that must be analyzed for video recovery. Changes to the values in this region resulting from deletion operations can have a direct impact on the feasibility of video recovery.

\section{Mechanisms of Honeywell NVR Video Deletion}
This section analyzes the internal mechanisms of three deletion methods provided by the Honeywell NVR device: formatting-based deletion, data expiration, and overwrite.



\subsection{Formatting-based Deletion}
\label{fdeletion}
Disk formatting deletes all stored video data and does not support partial deletion. As shown in \autoref{fig:FDeletion}(a)–(b), formatting modifies the GUIDs in the start sectors for both the disk and partitions, which in turn changes the header and partition table checksums. In Partition 1, the header is reset to its initial state after formatting, and all associated metadata (e.g., Block Index, Channel Index, and Record Index) are completely removed. As a result, metadata-based video retrieval becomes impossible, and all video-related information except the raw Video Data is erased, causing the NVR to recognize the disk as newly connected. 

When recording resumes after formatting-based Deletion (see \autoref{fig:FDeletion}(c)), new video data are written from the beginning of the video data region, progressively overwriting previously stored video data in real time (e.g., video data 1 is overwritten by video data 3). Video metadata is also rewritten starting from the beginning of each region.


\begin{figure*}[t]
  \centering
  \includegraphics[width=0.96\linewidth]{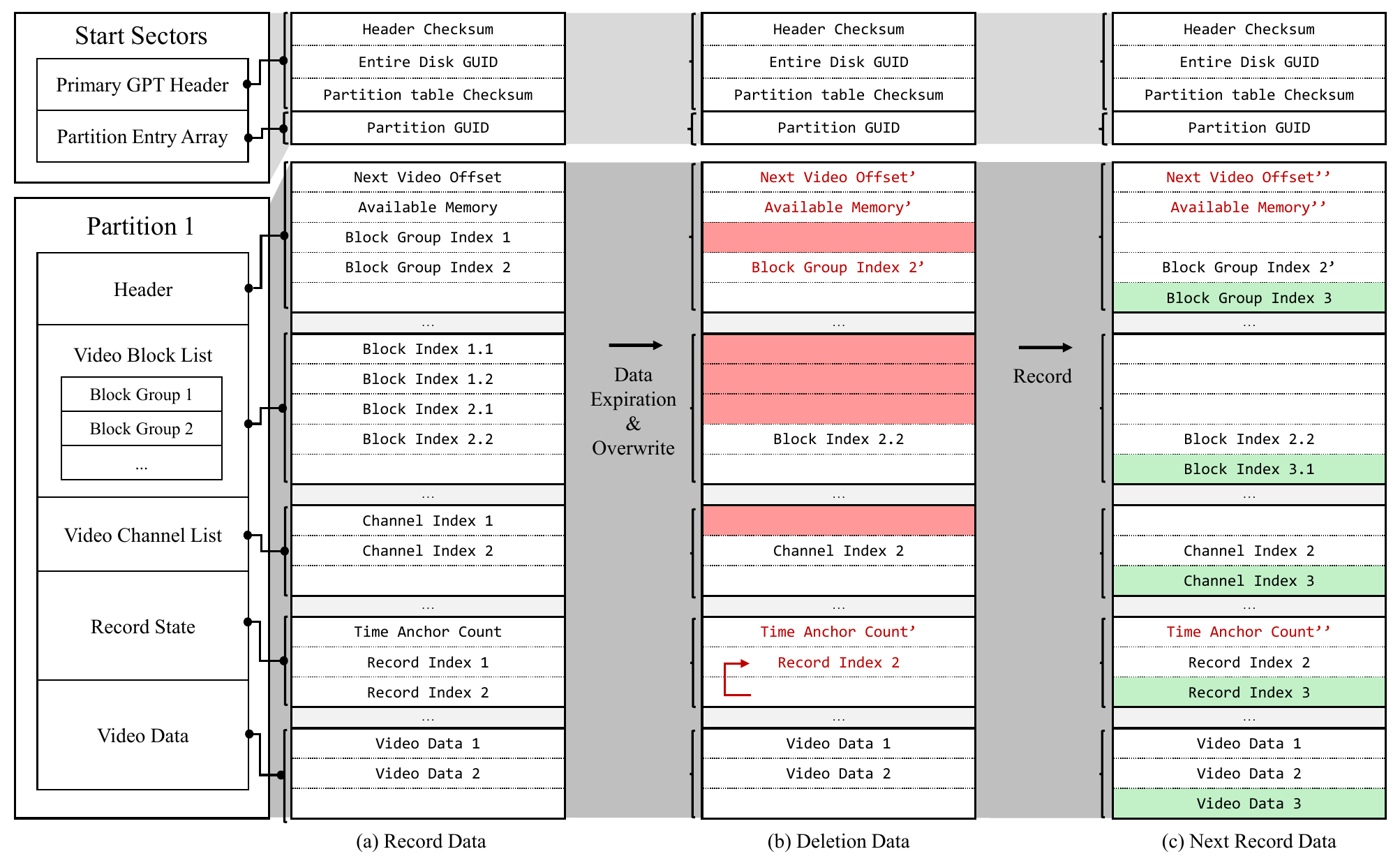}
  \caption{Changes in Start Sectors and Partition 1 caused by expiration \& overwrite deletion.}
  \label{fig:DExpiration}
\end{figure*}

\subsection{Data Expiration}
\label{edeletion}
When overwriting is disabled and an expiration period is configured, videos are deleted sequentially upon expiration, starting from the oldest recordings; selective deletion by time is not supported. 
As shown in \autoref{fig:DExpiration}(a)–(b), expiration-based deletion updates Partition 1 metadata. In the header, the available memory increases proportionally to the deleted data, and the block group start time---illustrated by the update to Block Group Index 2 in the figure---is updated to the oldest searchable recording time after the specified expiration period. 

In the Video Block List and Video Channel List, entries corresponding to expired videos are removed. If all blocks within a block group are deleted, the corresponding Block Group Index is removed from the header. If any block remains (e.g., Block Index 2.2), the block group is retained (e.g., Block Group Index 2) and its timestamp is updated to that of the oldest remaining block.
In the Record State area, Record Index entries corresponding to expired times are removed, and subsequent entries shift forward to fill the gap.

When recording resumes on the same disk (\autoref{fig:DExpiration}(c)), metadata are appended to the Video Block List and Video Channel List rather than rewritten from the beginning, and video data are written starting after the latest remaining data. Once Partition 1 is filled, recording continues by wrapping around to the beginning of the video data region.

\subsection{Overwrite}
\label{odeletion}
When the video expiration period is set to zero and overwriting is enabled, the system deletes the oldest videos once the disk is full. This method removes videos sequentially from the oldest and does not support selective deletion by time.
When this deletion method is applied, Partition 1 metadata are updated in the same manner as expiration-based deletion. The primary difference is that the available memory value remains unchanged or varies only minimally, as the oldest videos are deleted concurrently with new recordings. All other metadata are modified or removed in the same manner as in expiration-based deletion.

\begin{figure*}[t]
  \centering
  \includegraphics[width=\linewidth]{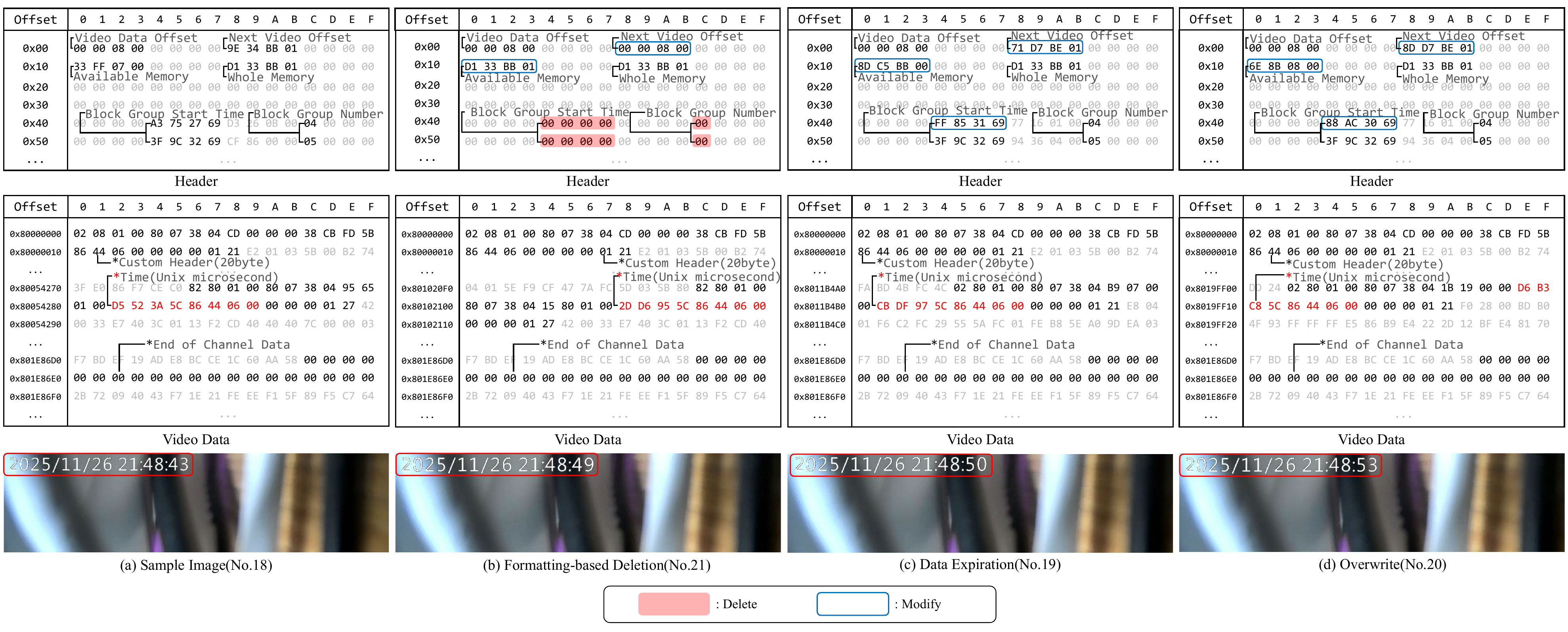}
  \caption{Header and Video Data region layouts under different deletion mechanisms—(a) normal recording, (b) formatting-based deletion, (c) expiration-based deletion, and (d) overwriting when storage is full—along with screenshots of successfully recovered video streams.}
  \label{fig:vrecovery}
\end{figure*}

\subsection{Implications of Analysis}
Our analysis yields the following implications regarding video deletion behavior:
\begin{itemize}
\item Data expiration and overwrite exhibit similar operational behaviors, whereas formatting-based deletion differs substantially.
\item None of the deletion methods explicitly erase video data in the Video Data region.
\item None of the three deletion methods support removal of video data at a specific point in time.
\item An attacker seeking to eliminate video from a particular moment is therefore most likely to use formatting-based deletion, which removes all recorded videos.
\item After formatting-based deletion, new recordings overwrite data from the beginning of the storage area, significantly reducing the likelihood that residual data remain.
\item In contrast, with data expiration and overwrite based deletion, new video data are written after existing data, increasing the likelihood that deleted video may persist depending on the time range.
\end{itemize}
Based on these observations, we examine the feasibility of video recovery under each deletion method.

\begin{table}[t]
\centering
\footnotesize
\caption{Comparison of deletion mechanisms in Honeywell NVR systems.}
\label{tab:deletion_comparison}
\resizebox{\columnwidth}{!}{
\begin{tabular}{lrrr}
\toprule
\textbf{Category} & \textbf{Formatting} & \textbf{Expiration} & \textbf{Overwrite} \\
\midrule
Deletion Scope 
& All data 
& Old data only 
& Old data only \\

Metadata Handling 
& Fully reset 
& Partially removed 
& Partially removed \\

Video data (Raw) 
& Remains 
& Remains 
& Remains \\

Write Pattern 
& Restart 
& Append
& Append \\

Recoverability 
& Medium 
& High 
& High \\

\bottomrule
\end{tabular}
}
\end{table}
\section{Case Study: Video Recovery}
This section analyzes the recoverability of deleted video data from Honeywell NVR devices. We use disk image \#18 from \autoref{tab:nvr_summary}, which contains approximately 45 hours of video recorded on a 160 GB hard disk between November 26, 2025 (21:48–22:18) and December 3, 2025 (21:18) to December 5, 2025 (18:22). \autoref{fig:vrecovery}(a) shows the Header and Video Data regions of image \#18, which include video data with the timestamp 0x000644865C3A52D5 (2025-11-26 21:48:43).
Using image \#18, we perform formatting-based (image \#21), expiration-based (image \#19), and overwrite-based (image \#20) deletions and conduct case studies on recovering the deleted video data.

Table~\ref{tab:deletion_comparison} summarizes the differences among the three deletion mechanisms. Detailed analysis of each mechanism is provided in the following subsections.


\subsection{Formatting-based Deletion}\label{frecovery}
\autoref{fig:vrecovery}(b) illustrates how the Header and Video Data regions change from \autoref{fig:vrecovery}(a) after formatting-based deletion. After deletion, the available memory becomes the total memory (0x01BB33D1), the video start offset is reset to the beginning of the video data region (0x80000000), and the Block Group Index data is completely erased, preventing recovery of deleted video data using Header information.

However, the Video Data region remains intact. The value 0x000644865C95D62D corresponds to the Unix microsecond timestamp 2025-11-26 21:48:49, indicating deleted data. We classify it as deleted video data because the Header region contains no video metadata (e.g., block group start times), yet video payloads persist in the Video Data region.
As described in \autoref{videodata}, video streams are separated by a 20-byte 0x00 delimiter, allowing direct extraction from the custom header containing deleted video's timestamp to the next delimiter. The extracted streams were saved as .dat files and successfully played using \texttt{ffplay}~\citep{ffplay}. Although the data includes a 20-byte Honeywell-specific Custom Header (see \autoref{fig:videoData}), \texttt{ffplay} decodes it correctly by skipping the customer header and processing the subsequent standard H.264 NAL units. Note that playback also succeeds when the Custom Header is removed from the extracted streams.

\begin{tcolorbox}[
  colback=white,
  colframe=black!70,
  boxrule=0.7pt,
  height=2.6cm
]
\textbf{Takeaway.}
Our experiments show that deleted video data can be recovered even when Header data is completely erased by formatting-based deletion. However, since new recordings overwrite the oldest data in the Video Data region, timely recovery is critical. 
\end{tcolorbox}




\subsection{Data Expiration}
\autoref{fig:vrecovery}(c) shows the Header and Video Data regions after expiration-based deletion. Unlike formatting-based deletion, the Header is not reset; instead, the available memory and next offset fields are updated to reflect overwriting of expired videos with new data. The block group start time is set to the timestamp of the oldest retained video (0x693185FF, corresponding to December 4, 2025, 13:00).
Deleted video data can be identified by comparing this start time with per-stream timestamps in the Video Data region. Streams with timestamps earlier than the Header start time are classified as deleted. For example, a stream with timestamp 0x000644865C97DFCB (2025-11-26 21:48:50) is confirmed as deleted.

Recovery proceeds identically to formatting-based deletion: deleted streams are extracted from the custom header containing the deleted video’s timestamp to the next delimiter and can be successfully played using \texttt{ffplay}. Recovery is more likely to be successful under expiration-based deletion, as video data in the Video Data region remains intact unless the storage is full.

\begin{tcolorbox}[
  colback=white,
  colframe=black!70,
  boxrule=0.7pt,
  height=3.0cm
]
\textbf{Takeaway.}
\WC{1} Under expiration-based deletion, video data typically remains intact unless the disk is full, resulting in a higher probability of successful recovery. \WC{2} Deleted video data can be identified by comparing per-stream timestamps with the Header Block Group Index start time.
\end{tcolorbox}


\subsection{Overwrite}
\autoref{fig:vrecovery}(d) shows the Header and Video Data regions after overwriting. This deletion mechanism follows expiration-based deletion; thus, the same recovery method applies. The key difference is that deletion occurs only when the disk is full, causing the oldest video data to be overwritten. 
Using the same recovery method as expiration-based deletion, we successfully recovered deleted video data.


\section{Discussion and Future Work}
This study analyzes the Honeywell NVR file system and evaluates the recovery potential of video data under different deletion mechanisms. However, in real-world scenarios, the recoverability may vary depending on recording conditions, system configurations, and elapsed time. 

In this study, we demonstrate that video recovery is feasible, and leave a comprehensive quantitative evaluation for future work. Specifically, tasks such as detecting formatting events, reconstructing deletion timelines, identifying deleted content (e.g., playback duration or channel), and distinguishing deletion types are left for future work. 

We believe our findings are significant for the forensic community, as this is the first study to uncover previously undocumented file system structures and internal deletion behaviors of Honeywell NVRs. Our work demonstrates that meaningful visual evidence can be recovered for criminal investigations.
However, our analysis is currently limited to a single Honeywell device. To improve generality, future work will evaluate additional models from the same manufacturer. Given that Honeywell NVRs are likely to share common file system designs and deletion behaviors, we expect the proposed recovery methods to generalize to other models with minimal modification. In addition, further reverse engineering of file system structures is expected to improve the reliability and scalability of our findings.





\section{Conclusion}
Real-time video surveillance systems rely on NVRs that store large volumes of video data using proprietary and undocumented file systems, posing significant challenges for forensic analysis. In this work, we present the first in-depth analysis of the proprietary file system used by Honeywell surveillance devices and examine its deletion mechanisms and implications for video recovery.
Using binary diffing and low-level file system analysis, we characterize three deletion methods---formatting-based deletion, data expiration, and overwriting---and analyze their effects on metadata and on-disk video data. Our results demonstrate that, despite deletion operations, meaningful video evidence can often be recovered directly from the Video Data region, even in the absence of conventional file system metadata.

\section*{Acknowledgements}
We thank the anonymous reviewers 
	for their constructive feedback.
This work was partly supported by the three
Institute of Information \& Communications 
Technology Planning \& Evaluation (IITP) 
grants funded by the Korean government 
(MSIT; Ministry of Science and ICT) 
(No.2024-00337703; Development of satellite security vulnerability detection techniques using AI and specification-based automation tools, No.2024- 00398745; Proofs and responses against evidence tampering in the new digital environment, No. 2022-0-00688; AI Platform to Fully Adapt and 
Reflect Privacy-Policy Changes, IITP-2025-RS-2024-00437849; ICT Challenge and Advanced Network of HRD).

\bibliographystyle{elsarticle-harv}
\bibliography{references}

\end{document}